\documentstyle[12pt]{article}

\begin{document}
\topmargin 0pt
\oddsidemargin 0mm
\vspace{3mm}
\vspace{10mm}
\begin{center}
{\Large{\bf Resonance Contributions to $\eta$
Photoproduction on Protons Found Using  
Dispersion Relations and an Isobar Model}}\\
\vspace{1cm}
{\large I.G.Aznauryan}\\
\vspace{1cm}
{\em Yerevan Physics Institute,
Alikhanian Brothers St.2, Yerevan, 375036 Armenia}\\
{(e-mail addresses:  aznaury@jlab.org, aznaur@jerewan1.yerphi.am)}\\
\vspace{5mm}
Abstract\\
\end{center}
\vspace{3mm}
The contributions of the resonances
$D_{13}(1520)$, $S_{11}(1535)$, $S_{11}(1650)$, $D_{15}(1675)$, 
$F_{15}(1680)$,
$D_{13}(1700)$, $P_{11}(1710)$, $P_{13}(1720)$
to $\gamma p\rightarrow \eta p$
are found from the data on cross sections, beam and target
asymmetries using two approaches: fixed-t dispersion
relations and an isobar model. Utilization of the two
approaches and comparison of the results obtained
with different parametrizations 
of the resonance contributions allowed us to make conclusions
about the model-dependence of these contributions.
We conclude that the results for the contributions
of the resonances $D_{13}(1520)$, $S_{11}(1535)$,  
$F_{15}(1680)$ to corresponding multipole amplitudes
are stable. With this the results for
$D_{13}(1520)$ and $F_{15}(1680)$,
combined with their PDG photoexcitation helicity
amplitudes,  allowed us to find the branching ratios
$Br \left(D_{13}(1520)\rightarrow \eta N\right)=0.05\pm 0.02~\%$,
$Br \left(F_{15}(1680)\rightarrow \eta N\right)=0.16\pm0.04~\%$
which have significantly better accuracy than the PDG data.
The total Breit-Wigner width
of the  $S_{11}(1535)$ is model-dependent, we have
obtained $\Gamma \left(S_{11}(1520)\right)=142~MeV$
and $195~MeV$
using dispersion relations and the isobar model, respectively.
The results for the $S_{11}(1650)$, $D_{15}(1675)$,
$P_{11}(1710)$,  $P_{13}(1720)$ are model dependent, only 
the signs and  orders of magnitude of their contributions
to multipole amplitudes are determined.
The results for the
$D_{13}(1700)$ 
are strongly model-dependent.
\newline
\vspace{1cm}
\vspace{10mm}
\renewcommand{\thefootnote}{\arabic {footnote}}
\setcounter{page}{1}
\section{Introduction}

It is well known that photo- and electroproduction
of $\eta$ on nucleons provide a unique opportunity
for detailed study of the properties of the
$S_{11}(1535)$ resonance, because this resonance has a large
branching ratio into the $\eta N$ channel, unlike
other resonances with close masses: 
$D_{13}(1520)$, $S_{11}(1650)$, $D_{15}(1675)$, $F_{15}(1680)$,
which overlap with the $S_{11}(1535)$
in $\pi$ photoproduction.
The situation with overlapping resonances is significantly
simplified in the case of  $\eta$ in comparison with  $\pi$
also for the reason that the resonances with isotopic spin
$3/2$ do not contribute to $\eta$ photoproduction.

Investigation of $\eta$ photoproduction
is interesting also for the following reasons.
Investigating the contributions of the resonances with
small $\eta N$ branching ratios
and well known  
photocouplings, we have the possibility of specifying
these  branching ratios.
Eta photoproduction
provides also the possibility of searching for 
the "missing resonances", which cannot be observed
in $\pi N$ scattering and  $\pi$ photoproduction
on nucleons.

The approaches which are used for extraction of resonance
properties from experimental data  
can be divided into two groups. 
One group is based on  coupled channels calculations, 
which has been  used mostly
to analyse $\pi N$ data  \cite{1,2,3}.
The other group  consists of the approaches
which extract resonance properties from
single reaction data by modeling background
and parametrizing  the 
resonance contributions to
multipole amplitudes according to the
Breit-Wigner formula.
For the $\pi N$ scattering
such an analysis is made in Ref. \cite{4}.
For the pion
photo- and electroproduction on the nucleons
the latest analyses of this type are done
in Refs. \cite{5,6}.

In the case of $\eta$
photoproduction on protons, there is an analysis
made within  the coupled channel
approach based on  effective Lagrangians \cite{7}.
At energies above $1.54~GeV$ only sparse data
on the differential cross sections were used in this analysis.
More recent data on the differential cross sections
from GRAAL \cite{8} 
and CLAS \cite{9}, and
the polarized beam asymmetry
from GRAAL \cite {10}  were not available at that time. 
A more complete data set \cite{8,10,11,12}
was analysed in Refs. \cite{13,14} 
using isobar models 
with parametrization of the resonance contributions
according to the Breit-Wigner formula.
The background in Ref. \cite {13}
is built from  $s$ and $u$ channel nucleon
contributions (the Born term) and  $t$ channel
vector meson exchanges. The model of Ref. \cite{14}
is based
directly on the utilization of the quark model results.
$t$ channel
vector meson exchanges are excluded in this model
in order to avoid  possible
double counting of contributions from
the $s$ and $t$ channels;
the background 
is built from the $u$ channel resonance contributions.

In this paper we will investigate the data on 
$\eta$ photoproduction on protons from Refs. \cite{8,9,10,11,12} 
using fixed-$t$ dispersion
relations for invariant amplitudes. The imaginary parts of the amplitudes we will
build, as in isobar models, from the $s$ channel   
resonance contributions parametrized in 
the Breit-Wigner form. Using dispersion relations
we will find the real parts of the amplitudes, which 
include the contributions of the nucleon
poles in the  $s$ and $u$ channels (the Born term).
They include also the integrals
over the imaginary parts of the resonance contributions.
As a result, in addition to the $s$ channel contributions,
$u$ channel resonance contributions to
the real parts of the amplitudes 
will be reproduced  due to the crossing symmetry. 
Also, the point connected with the  $t$ channel
vector meson contributions
does not arise in the dispersion relations approach.
They do not  directly enter
the dispersion relations. Such contributions could be imitated
by high energy contributions to the dispersion integrals.
However, our estimations made for pion photoproduction
on nucleons, where we have enough information for the estimation
of these contributions, show that their role in the description
of the data in the second and third resonance regions is negligibly
small \cite{15}. The dispersion relations approach will
be presented in Sec. 2.
In this paper we will present also our results on the description
of the same data set in an isobar model
which is very close to the model of Ref. \cite {13}.
The difference lies only in the slightly
different parametrization of the resonance contributions
and in the inclusion in our analysis
of the Roper resonance $P_{11}(1440)$.
This approach will be presented in Sec. 3.
Comparison of the results obtained in the two approaches
using the same data set
will allow us to make conclusions on the model-dependence
of the extracted resonance characteristics.
We will compare also with the results
obtained in  Ref. \cite {13} using an isobar model. This will allow us
to check the dependence on the details
of the model, and on the observables
used in the analysis.
The results obtained will be presented and discussed in Sec. 4.
The conclusions are presented in Sec. 5.
    
\section{Dispersion relations}

In the dispersion relations approach we use 
fixed-t dispersion relations for 
invariant amplitudes.
All formulas in this paper we will write
for the $\eta$ electroproduction on the nucleons, i.e.
for the reaction $\gamma^* N\rightarrow \eta N$.
The amplitudes of this reaction
we choose following the work  \cite{16}
in accordance with the definition of the hadron 
electromagnetic current in the form:
\begin{eqnarray}
I^\mu =\bar{u}(p_2)\gamma _5 \left\{ \frac{B_1}{2}\left[ \gamma
^\mu (\gamma \tilde{k})-(\gamma \tilde{k})\gamma ^\mu \right]+2P ^\mu
B_2+2\tilde{q}^\mu B_3+2\tilde{k}^\mu B_4\right.\nonumber \\
\left.-\gamma ^\mu B_5+(\gamma \tilde{k})P^\mu B_6+
(\gamma \tilde{k})\tilde{k}^\mu B_7+
(\gamma \tilde{k})\tilde{q}^\mu B_8\right\} u(p_1),
\label{1}\end{eqnarray}
where $\tilde{k},~\tilde{q},~p_1,~p_2$ are 4-momenta
of the virtual photon, $\eta$, initial and final nucleons, respectively, 
$P=\frac{1}{2}(p_1+p_2),~B_1,B_2,...B_8$
are invariant amplitudes which are functions
of the invariant variables 
$s=(\tilde{k}+p_1)^2,~t=(\tilde{k}-\tilde{q})^2,~Q^2\equiv 
-\tilde{k}^2$.

The conservation of the hadron electromagnetic current
leads to the relations:
\begin{eqnarray}
&&4Q^2B_4=(s-u)B_2 -2(t+Q^2-m_{\eta} ^2)B_3, \nonumber\\
&&2Q^2B_7=-2B'_5 -(t+Q^2-m_{\eta} ^2)B_8,
\label{2}\end{eqnarray}
where
$B'_5\equiv B_5-\frac{1}{4}(s-u)B_6$.
So, only the six of the eight invariant amplitudes
are independent. 
As independent amplitudes let us choose
$B_1,B_2,B_3,B'_5,B_6,B_8$.
For all these amplitudes 
unsubtracted dispersion relations at fixed $t$ can be written.
The relations between 
$B_1,B_2,B_3,B'_5,B_6,B_8$
and the multipole amplitudes and observables
are given in Appendix A. 

Unlike the $\pi$ electroproduction,
in the case of $\eta$, dispersion relations
for $p$ and $n$ are independent from each other
and can be written separately. By this reason we write
dispersion relations for $N$, supposing that $N=p$ or $n$:
\begin{eqnarray}
Re~ B^N_i(s,t,Q^2)=&&eg_{\eta NN}R^N_i(Q^2)
\left(\frac{1}{s-m_N^2}+ 
\frac{\eta_i}{u-m_N^2}\right)\nonumber\\
&&+\frac{P}{\pi }\int \limits_{s_{cut}}^{\infty}
Im~ B^N_i(s',t,Q^2)
\left(\frac{1}{s'-s}+ \frac{\eta_i }{s'-u}\right) ds',
\label{3}\end{eqnarray}
where
$R^N_i(Q^2)$ correspond to the residues in 
the nucleon poles:
\begin{eqnarray}
&&R^N_1(Q^2)=F^N_1(Q^2)
+2m_NF^N_2(Q^2),\nonumber \\
&&R^N_2(Q^2)=-F^N_1(Q^2),\nonumber \\
&&R^N_3(Q^2)=-\frac{1}{2}F^N_1(Q^2), \\
&&{R_5^N}'(Q^2)=\frac
{m_{\eta}^2 -Q^2-t}{2}F^N_2(Q^2),\nonumber \\
&&R^N_6(Q^2)=2 F^N_2(Q^2),\nonumber \\
&&R^N_8(Q^2)=F^N_2(Q^2),\nonumber
\end{eqnarray}
$e^2/4\pi=1/137$, $g_{\eta NN}$ is the $\eta NN$   
pseudoscalar coupling constant, and
$F_1^{N}(Q^2),~F_2^{N}(Q^2)$ are the nucleon Pauli
form factors. In the case of the real photons
these form factors are normalized to $F_1^{p}(0)=1$, $F_1^{n}(0)=0$,
$F_2^{N}(0)=
\frac{\kappa_N}{2m_N}$,
$\kappa_p=1.79$, $\kappa_n=-1.91$.

The factors $\eta_i$ in the 
dispersion relations define the 
crossing symmetry properties of 
the invariant amplitudes, they 
are equal to $\eta_1=\eta_2=\eta_6=1,~\eta_3=\eta'_5=\eta_8=-1$. 

Other distinctive feature of the dispersion relations
for $\eta$ electroproduction is the presence
in the dispersion integrals of the unphysical region
from $s=s_{cut}=(m_N+m_{\pi^0})^2$ to $s=s_{thr}=(m_N+m_{\eta})^2$.
In this region we take into account the contribution
of the Roper resonance $P_{11}(1440)$. Also, we continue
to this region the contributions of $S_{11}(1535)$
and $S_{11}(1650)$ in the way which will be described below.
In the unphisical
region of the dispersion integrals, 
we do not take into account 
contributions of other resonances with higher masses
as they are strongly
suppressed in comparison with the contribution
of $S_{11}(1535)$.

The contribution of the Roper resonance $P_{11}(1440)$ to  dispersion
integrals was found by evaluating the Feynman diagrams
for  $\gamma^*+N\rightarrow R \rightarrow \eta+N$
and the corresponding invariant
amplitudes using   the effective Lagrangians:

\begin{eqnarray}
&&{\cal L}^{(1)}_{\gamma NR}=-eF_1^{R}(Q^2)\bar{\psi}_{R}
\left[({\partial}_{\mu}{\partial}^{\mu})(\gamma_{\nu}A^{\nu})
-i(m_{R}-m_N){\partial}_{\mu}A^{\mu}\right]\psi_N,\nonumber \\
&&{\cal L}^{(2)}_{\gamma NR}=-eF_2^{R}(Q^2)\bar{\psi}_{R}
(\sigma_{\mu\nu}{\partial}^{\mu}A^{\nu})\psi_N, \\
&&{\cal L}_{\eta NR}=-ig_{\eta NR}\bar{\psi}_{R}
\gamma_5\psi_N\phi_{\eta},\nonumber 
\end{eqnarray}
with $A^{\mu}$ the electromagnetic field,
and $\psi_N,~\psi_{R},~\phi_{\eta}$ the field operators
of the nucleon, $P_{11}(1440)$ and $\eta$.
With this, the $s$-channel $P_{11}(1440)$ contributions
into $B_i(s,t,Q^2)$ have the following form:
\begin{equation}
B_i^R(s,t,Q^2)=eg_{\eta NR}R^R_i(Q^2)
\frac{1}{s-m_{R}^2+im_{R}\Gamma_{tot}},
\end{equation}
where
\begin{eqnarray}
&&R^R_1(Q^2)=Q^2F^R_1(Q^2)
+(m_N+m_{R})F^R_2(Q^2),\nonumber \\
&&R^R_2(Q^2)=-Q^2F_1^{R}(Q^2),\nonumber \\
&&R^R_3(Q^2)=-\frac{Q^2}{2}F_1^{R}(Q^2), \\
&&{R^R_5}'(Q^2)=(m_{R}-m_{N})Q^2F_1^{R}(Q^2)+
\frac{m_{\eta}^2 -Q^2-t}{2}F_2^{R}(Q^2),\nonumber \\
&&R^R_6(Q^2)=2F_2^{R}(Q^2),\nonumber \\    
&&R^R_8(Q^2)=F_2^{R}(Q^2).\nonumber
\end{eqnarray}
$\Gamma_{tot}$ we parametrize according to
the formulas presented below, taking the mass and the width
of $P_{11}(1440)$ equal to $M=1440~MeV$, $\Gamma=350~MeV$,
and the branching ratio into the $\pi N$ channel $\beta_{\pi N}=0.6$.
The form factor $F_2^{R}(Q^2)$ at $Q^2=0$ is related to the
$P_{11}(1440)\rightarrow N\gamma$ helicity amplitudes:
\begin{equation}
A_{1/2}=e\left(\pi k\frac{m_{R}}{m_N}\right)^{1/2}F_2^{R}(0),
\end{equation} 
and can be found from the Particle Data Group \cite{17} values for $A_{1/2}$:
\begin{equation}
F_{2}^{R}(0)\equiv\frac{\kappa^R_N}{2m_N},~ \kappa^R_p=-0.5,
~\kappa^R_n=0.3.
\end{equation} 

The  $g_{\eta NN}$ and  $g_{\eta NR}$ coupling constants
we will found from the analysis of the $\eta$ photoproduction data.

Let us turn now to the imaginary parts of the amplitudes,
which we construct through the contributions of the resonances
$D_{13}(1520)$, $S_{11}(1535)$, $S_{11}(1650)$, $D_{15}(1675)$, $F_{15}(1680)$,
$D_{13}(1700)$, $P_{11}(1710)$, $P_{13}(1720)$.
The contributions of these resonances into
multipole amplitudes we parametrize using the Breit-Wigner
formula in the form based on the parametrizations
used in Refs. (\cite{13,16,18}):
\begin{equation}
{\cal M}(W,Q^2)={\cal M}(W=M,Q^2)\left(\frac{k}{k_r}\right)^n
\left(\frac{q_{\eta,r} }{q_{\eta}}\frac{k_r}{k}\frac{\Gamma_\eta \Gamma_\gamma }
{\beta_{\eta N} \Gamma^2 }\right)^{1/2}
\frac{M\Gamma}{M^2-W^2-iM\Gamma_{total}},
\end{equation}
where $n=0$ for $M_{l\pm},~E_{l\pm}$,  $n=1$ for $S_{l\pm}$ and
\begin{eqnarray}
  &&\Gamma_{total}=\Gamma_{\pi}+\Gamma_{\eta}+\Gamma_{inel},\\
 &&\Gamma_{\pi}=\beta_{\pi N}\Gamma\left(\frac{q_{\pi}}{q_{\pi,r}}\right)^{2l+1}
 \left(\frac{X^2+q_{\pi,r}^2}{X^2+q_{\pi}^2}\right),\\
 &&\Gamma_{\eta}=\beta_{\eta N}
\Gamma\left(\frac{q_{\eta}}{q_{\eta,r}}\right)^{2l+1}
 \left(\frac{X^2+q_{\eta,r}^2}{X^2+q_{\eta}^2}\right),\\
 &&\Gamma_{\gamma}=\Gamma\left(\frac{k}{k_r}\right)^{2l^{\prime}+1}
 \left(\frac{X^2+k_r^2}{X^2+k^2}\right)^{l^{\prime}},\\
 &&\Gamma_{inel}=
(1-\beta_{\pi N}-\beta_{\eta N})
\Gamma\left(\frac{q_{2\pi}}{q_{2\pi,r}}\right)^{2l+4}
 \left(\frac{X^2+q_{2\pi,r}^2}{X^2+q_{2\pi}^2}\right)^{l+2}.
\end{eqnarray}
For $M_{l\pm},E_{l+},S_{l+}$, $l'=l$;
for $E_{l-},S_{l-}$, $l'=l-2$ if $l\geq 2$;
for $S_{1-}$, $l'=1$;
$M$ and $\Gamma$ are masses and widths of the resonances,
$\beta_{\pi N}$ and $\beta_{\eta N}$ are their
branching ratios into the $\pi N$ 
and $\eta N$ channels,
$k$, $q_{\eta}\equiv q$, $q_{\pi}$ and $q_{2\pi}$ are the 3-momenta of 
the $\gamma$, $\eta$, $\pi$ and $2\pi$ system
in the decays of the resonances into the
 $\gamma N$, $\eta N$, $\pi N$ and $2\pi N$ channels in the c.m.s.,
$k_r$, $q_{\eta,r}$, $q_{\pi,r}$, $q_{2\pi,r}$ are the magnitudes
of these momenta at the resonance peak ($W=M$);
$X$ are phenomenological parameters, assumed
to be $500~MeV$ for all resonances, except $S_{11}(1535)$
and  $S_{11}(1650)$. For these resonances parameters
X were found by fitting experimental data.

Below the thresholds
of $2\pi+N$ and  $\eta+N$ productions we take,
respectively, $q_{2\pi}=0$ and $q_{\eta}=0$.
So, below $\eta N$ production threshold, 
$\Gamma_{total}=\Gamma_{\pi}+\Gamma_{inel}$,
and, respectively, below $2\pi N$  production
threshold, $\Gamma_{total}=\Gamma_{\pi}$.

Using the parametrizations (10-15) for the contributions of the resonances
$S_{11}(1535)$, $S_{11}(1650)$ to $E_{0+}$ and $S_{0+}$ , one can easily
continue their contributions to the invariant amplitudes
to the unphysical region ($s<(m_{\eta}+m_N)^2$)
of the dispersion integrals via the formulas:
\begin{eqnarray}
&&{\cal E}_{0+}\equiv \frac{E_{0+}}{2W[(E_1+m_N)(E_2+m_N)]^{1/2}},
~{\cal S}_{0+}\equiv \frac{Q^2S_{0+}}
{2W[(E_1-m_N)(E_2+m_N)]^{1/2}},\nonumber \\
&&B_1(s,t,Q^2)={\cal E}_{0+},\nonumber \\
&&B_2(s,t,Q^2)=2B_3(s,t,Q^2)=\frac{-Q^2(E_1+m_N){\cal E}_{0+}
+(W-m_N){\cal S}_{0+}}{2Wk^2}, \\ 
&&B_5(s,t,Q^2)=-(W+m_N){\cal E}_{0+},\nonumber \\
&&B_6(s,t,Q^2)=2B_8(s,t,Q^2)=-\frac{(W-m_N)(E_1+m_N){\cal E}_{0+}
+{\cal S}_{0+}}{Wk^2}.\nonumber  
\end{eqnarray}

It is seen that there are no irregularities
in the continuation of Eqs. (16) from $s\geq(m_{\eta}+m_N)^2$
to $s<(m_{\eta}+m_N)^2$.

\section{Isobar model}

The isobar model we use is very close to the model
of Ref. \cite{13}. It contains  contributions of  resonances,
and the nonresonance background build from the Born term
and the $t$-channel $\rho$
and $\omega$ contributions. 
Unlike the model
of Ref. \cite{13}, we take into account not only
the resonances with the masses above $\eta N$ production threshold,
but also the contribution of the Roper
resonance, which, in fact, can be considered
as the background contribution. This resonance
is introduced  in a form described
in the previous Section. 
There is also a difference between our parametrizations
of the resonance contributions which consists in the extra factor
$W_R/W$ in the expressions for $\Gamma_{\pi}$
and $\Gamma_{\eta}$ of  Ref. \cite{13}, 
and in the factor $\Gamma_{\gamma}/\Gamma$ in the parentheses
of Eq. (10). 

We have found the  $t$-channel
$\rho$ and $\omega$ 
exchange contributions  
to  $\gamma^*+N\rightarrow \eta+N$ 
by evaluating the Feynman diagrams
for this process
and the corresponding invariant
amplitudes using   the effective Lagrangians:

\begin{eqnarray}
&&{\cal L}_{\gamma \eta V}=\frac{e\lambda_V}{m_{\eta}}F^{(V)}(Q^2)
\varepsilon_{\mu\nu\sigma\rho}
{\partial}^{\mu}A^{\nu}
{\partial}^{\sigma}V^{\rho}\phi_{\eta}, \\
&&{\cal L}_{VNN}=\bar{\psi}_{N}
\left(g_{1V}\gamma_{\mu}+\frac{g_{2V}}{2m_N}
\sigma_{\mu\nu}{\partial}^{\nu}\right)V^{\mu}\psi_N. 
\end{eqnarray}
The electromagnetic coupling constants $\lambda_V$
are related to the $V\rightarrow \eta\gamma$ 
radiative decay widths via 
\begin{equation}
\Gamma_{V\rightarrow \eta\gamma}=
\frac{e^2 q_{\eta}^3}{12\pi m_{\eta}^2}\lambda_V^2,
\end{equation}
where $q_{\eta}$ is the $\eta$ momentum
in this decay.
The values of $\lambda_V$ found using PDG data \cite{17}
are presented in Table \ref{tab1}.
The form factors $F^{(V)}(Q^2)$ are introduced to describe 
the $Q^2$ dependence of
the $\gamma \eta V$ couplings, at $Q^2=0$ they are normalized
to $F^{(V)}(0)=1$.
The off-shell behaviour of the hadronic couplings
were described according to Ref. \cite{13} in the form
\begin{equation}
g_{iV}=\tilde{g}_{iV}\left(\frac{\Lambda_V^2-m_V^2}
{\Lambda_V^2-t}\right)^2,~i=1,2,                         
\end{equation}
with $\Lambda_V=1.3~GeV$.
The contributions of $\rho$ and $\omega$ exchanges
to the invariant amplitudes are:
\begin{eqnarray}
&&B_{1,V}=\frac{e\lambda _V}{m_\eta}
\left[2m_Ng_{1V }+t\frac{g_{2V}}
{2m_N}\right]\frac{1}{t-m_{V}^2},\nonumber\\
&&B_{2,V}=\frac{e\lambda _V}{m_\eta}
\frac{g_{2V}}{4m_N}
\frac{Q^2+m_{\eta}^2-t}{t-m_{V}^2},\\
&&B_{3,V}=\frac{e\lambda _V}{m_\eta}
\frac{g_{2V}}
{8m_N}\frac{u-s}{t-m_{V}^2},~~   
B_{6,V}=2\frac{e\lambda _V}
{m_\eta}\frac{g_{1V}}{t-m_V^2}.\nonumber
\end{eqnarray}

The coupling constants $\tilde{g}_{iV}$
are not well known, their ranges 
from Ref. \cite{19} are presented
in Table 1. In our analysis, we consider them as free parameters.
The values of $\tilde{g}_{iV}$ found from the fitting
of data on the $\eta$ photoproduction on the protons
within the model presented in this Section are given
in Table 1.

\section{Results and discussion}

Using the two approaches described in Sections 2 and 3, we have fitted
the experimental data on the $\eta$ photoproduction on protons.
We have used the information on the differential cross sections 
from TAPS \cite{11} ($W=1.491-1.537~GeV$, $E_{\gamma}=716-790~MeV$), 
GRAAL \cite{8} ($W=1.49-1.716~GeV$, $E_{\gamma}=714-1100~MeV$), 
and CLAS \cite{9} ($W=1.528-2.12~GeV$, $E_{\gamma}=775-1925~MeV$),
on the polarized beam asymmetry from GRAAL \cite {10} ($W=1.506-1.688~GeV$, 
$E_{\gamma}=740-1050~MeV$),
and  on the polarized target asymmetry from ELSA \cite {12}
($W=1.492-1.719~GeV$, $E_{\gamma}=717-1105~MeV$). 
As the role of the data, obtained with
the polarized beam and target, appeared to be very important,
we have restricted our investigation
by the energies $W=1.49-1.73~GeV$, where such data
are present. 

The fitted parameters in the dispersion relations
approach were the coupling constants
$g_{\eta NN}$, $g_{\eta NR}$, and the magnitudes
of the multipole amplitudes corresponding to the resonance
contributions at the resonance positions.
In the isobar model approach there are four additional
fitted parameters connected with the 
the $\rho$ and $\omega$ contributions: ${\tilde{g}}_{1\rho}$,
${\tilde{g}}_{2\rho}$, ${\tilde{g}}_{1\omega}$ and ${\tilde{g}}_{2\omega}$.
We have fitted also the masses of all resonances
taking into account the ranges given by
the PDG \cite{17}. The widths and parameters X of the most
prominent resonances: $S_{11}(1535)$ and  $S_{11}(1650)$,
were fitted too. The widths of other
resonances used in our analysis 
are presented in Table 2.
The used branching ratios correspond to the mean
values of the PDG data presented in this Table. 

In Table 3 we present the obtained 
values of the multipole amplitudes for the resonance
contributions at the resonance
positions. 
For comparison, multipole amplitudes extracted from 
the results of Ref. \cite{13} are given.
We present also the 
photoexcitation helicity amplitudes
obtained using our results, and
the ranges for these amplitudes
from  the PDG data \cite{17}.

In Figs. 1-4 we present our results obtained using
both approaches in comparison with the experimental data.
The overall $\chi^2/datapoint$ was 1.24 and 1.22 
in the dispersion relations
approach and in the isobar model, respectively.

In Fig. 5 multipole amplitudes are presented.

From Figs. 1-4 it is seen that our results are in very
good agreement with the experimental data. The only case,
when the experimental data are described not so well,
is connected with
the target asymmetry at small energies: $E_{\gamma}=717,~738~MeV$,
where the data change sign near $90^{\circ}$. Such structure  
is not seen in both our approaches, it is not described
also in other models \cite{7,13,14}. 

$\underline{Resonance~
contributions~ to~ the~ multipole~ amplitudes}$.
The experimental data on the differential
cross sections obviously indicate the dominant
role of the $s$-wave
in the $S_{11}(1535)$ resonance region.
These data can be described by the single resonance
fit taking into account
only the contribution of this resonance.
With increasing energy, for good description
of the data it is necessary to introduce quite large contribution
of the $S_{11}(1650)$ resonance,  although its presence
visually is seen neither in the experimental data, nor in the 
multipole amplitude $E_{0+}$ obtained in our
analyses (Fig. 5). We have fitted the masses and
widths of the $S_{11}(1535)$ and $S_{11}(1650)$, the obtained values
are presented in Table 2. It is seen that the
Breit-Wigner width of the $S_{11}(1535)$
is model-dependent.
The obtained widths of the $S_{11}(1650)$
are smaller than the values given by the PDG. Let us note that 
the same result was obtained in Ref. \cite{13}.
From our results obtained using two approaches
and from comparison with the results of Ref. \cite{13}
it is seen that the contribution of the 
resonance $S_{11}(1535)$
to the multipole amplitude $E_{0+}$
is determined very well. It practically does not depend on the used approach
and on the difference in the parametrizations of the resonance
contributions used in this paper and in Ref. \cite{13}.
The results for  the  $S_{11}(1650)$ contribution
to $E_{0+}$ are also quite stable. The
sign and order of magnitude of this contribution are determined,
however, its absolute value
is somewhat different in different approaches.

In the  $S_{11}(1535)$
resonance region, 
where we have $s$-wave dominance,
the polarized beam asymmetry is determined
by the interference of the $E_{0+}$
and $E_{2-}+M_{2-}$ multipoles only \cite{13}:
\begin{equation}
\Sigma=3~sin^2\theta~Re[E^*_{0+}(E_{2-}+M_{2-})]/|E_{0+}|^2.
\end{equation}
By this reason,
having the beam asymmetry data, we are able to find
with good accuracy the $D_{13}(1520)$ resonance
contribution, although by itself it is very small.
Indeed, as it is seen from Table 3, the results
for  $D_{13}(1520)$ obtained using different approaches
are in good agreement with each other.

With increasing energy the beam asymmetry data acquire
a forward-backward asymmetry behaviour, which is very sensitive
to the $F_{15}(1680)$ contribution \cite{13,20}.
In order to demonstrate this, we have presented in Fig. 2
by the dotted curve
the results for $\Sigma$ at $E_{\gamma}=1050~MeV$
(in the center of the $F_{15}(1680)$)
obtained in the isobar model with the zero
$F_{15}(1680)$ contribution.
The sensitivity of $\Sigma$ to
the $F_{15}(1680)$ contribution makes it clear, why the values
of the $F_{15}(1680)$ multipoles
obtained in both approaches are in good agreement
with each other.
We found that the difference with 
the results of Ref. \cite{13} is caused by our
inclusion into the fitting procedure of the target asymmetry
data, which do not present in the analysis
of Ref. \cite{13}. 
 
The contributions of other resonances $D_{15}(1675)$, $D_{13}(1700)$,
$P_{11}(1710)$, $P_{13}(1720)$ are not related immediately
to some specific features in the behaviour of the observables
$\sigma$, $\Sigma$ and $T$. The results obtained in our
analyses using the two approaches, and the comparison with
the results of  Ref. \cite{13} allow us to make conclusion
about the reliability of the obtained 
$D_{15}(1675)$, $D_{13}(1700)$,
$P_{11}(1710)$, $P_{13}(1720)$ contributions to the multipole
amplitudes.
We have performed also the fitting of the experimental data
taking exactly the model of Ref. \cite{13}. It appeared, that
the inclusion of the Roper contribution 
practically does not affect the obtained results,
it leads only to a better description of the data with
better $\chi^2$. However, the inclusion
into the analysis of the T data appeared to be significant.
As a result, the conclusion which we make is following.
The results for the $D_{15}(1675)$,
$P_{11}(1710)$, and $P_{13}(1720)$
do not depend strongly on the used 
approach. 
The signs and  orders of  magnitude of
the contributions of these resonances
are determined.
The strong difference with \cite{13} is connected
with the inclusion 
of the T data
into our analysis. 
The results for  the $D_{13}(1700)$ 
strongly depend on the model. Even the masses
of this resonance, obtained in two approaches, strongly differ
from each other (Table 2).

$\underline{Photoexcitation~ helicity~ amplitudes}$.
These amplitudes were calculated using the multipole amplitudes
presented in Table 3 by the formulas:
\begin{eqnarray}
&&\tilde{\cal M}_{l\pm}\equiv{\cal M}_{l\pm}\zeta_{\eta N}
\left[(2J+1)\pi\frac{q_{\eta,r}}{k_r}\frac{M}{m_N}
\frac{\Gamma_{tot}^2}{\Gamma_{\eta N}}\right]^{1/2},\\
&&A_{1/2}^{l+}=\frac{1}{2}\left[(l+2)\tilde{E}_{l+}+l\tilde{M}_{l+}\right],\nonumber \\
&&A_{3/2}^{l+}=\frac{\left[l(l+2)\right]^{1/2}}{2}
(\tilde{M}_{l+}-\tilde{E}_{l+})\\
&&A_{1/2}^{(l+1)-}=\frac{1}{2}\left[l\tilde{E}_{(l+1)-}
-(l+2)\tilde{M}_{(l+1)-}\right],\nonumber \\
&&A_{3/2}^{(l+1)-}=\frac{\left[l(l+2)\right]^{1/2}}{2}
(\tilde{E}_{(l+1)-}+\tilde{M}_{(l+1)-}),\nonumber 
\end{eqnarray}
where $J$ is the spin of the resonance, and $\zeta_{\eta N}$
is the relative sign between the $NN\eta$ coupling constant
in the Born term and 
the $N^*N\eta$ coupling constant,
which enter the relation (23)
through $\zeta_{\eta N}~\Gamma^{1/2}_{\eta N}\sim g_{N^*N\eta} $. 
The signs which we have used
are presented in Table 3. 
For the calculation of the photoexcitation helicity amplitudes
we have used the masses, widths, and branching ratios given 
in Table 2.
All used branching ratios are within the
ranges given by the PDG, except  the $P_{13}(1720)$. 
Here we had to take larger $\beta_{\eta N}$ in order
to stay within the ranges for ${}_pA_{1/2}$ given by the PDG.     

In the case of the $S_{11}(1535)$, there is
the difference between the values of the  photoexcitation helicity amplitude
${}_pA_{1/2}$ obtained using  dispersion relations and the isobar model.
It is connected with the different values
of masses and widths obtained in these approaches.
Both values are in good agreement
with the results obtained in other analyses
of the $\eta$ photoproduction: $95\div 140~\mu kb^{1/2}$ \cite{17}.
As in all these analyses our ${}_pA_{1/2}$ amplitudes
for the $S_{11}(1535)\rightarrow p\gamma$ transition
are larger than   the values extracted from 
$\pi$ photoproduction data.
However, the extraction of the amplitude  ${}_pA_{1/2}$
from the multipole amplitudes $E_{0+}$
is connected with the uncertainties caused by the
mass, width, and branching ratios of the $S_{11}(1535)$.
By this reason, for the comparison of the results
obtained in the $\pi$ and  $\eta$ photoproduction,
it is more reasonable to compare directly the results for 
the $S_{11}(1535)$ contributions to $E_{0+}$.
They are connected by the relation:
\begin{equation}  
E^{res}_{0+}(\gamma p\rightarrow \eta p)=E^{res}_{0+}(\gamma p\rightarrow \pi p)
\left(\frac{q_{\pi,r}}{q_{\eta,r}}\right)^{1/2}
\left(\frac{\beta_{\eta N}}{\beta_{\pi N}}\right)^{1/2},
\end{equation}  
which does not contain the uncertainty caused by the width
of the $S_{11}(1535)$. From the PDG data presented in Table 2
we have:
\begin{equation}
\left(\frac{q_{\pi,r}}{q_{\eta,r}}\right)^{1/2}=1.47\div 1.71,
\left(\frac{\beta_{\eta N}}{\beta_{\pi N}}\right)^{1/2}=0.8\div 1.35.
\end{equation}  
So, the maximum value of $E^{res}_{0+}(\gamma p\rightarrow \eta p)$
which we can obtain from Eq. (25) is
\begin{equation}  
E_{0+}^{res,max}(\gamma p\rightarrow \eta p)
=2.3~E^{res}_{0+}(\gamma p\rightarrow \pi p).
\end{equation}  
The magnitude of  $E^{res}_{0+}(\gamma p\rightarrow \pi p)$,
extracted from the results of Ref. \cite{5} 
using the Breit-Wigner formula for the
$S_{11}(1535)$ and $S_{11}(1650)$ contributions, is by our estimations 
$<0.6~\mu kb^{1/2}$. This gives 
$E_{0+}^{res,max}(\gamma p\rightarrow \eta p)<1.4~\mu kb^{1/2}$,
which is much smaller than the values in Table 3.
Therefore, indeed, there is disagreement between 
the $S_{11}(1535)$ contributions into $E_{0+}$
extracted from the $\pi$ and  $\eta$ photoproduction.
It is possible, that this disagreement is connected with the fact
that we compare the values
which correspond to the so-called "dressed" verteces
$\gamma p S_{11}(1535)$, i.e.
they contain the contributions  
caused by the rescattering effects in the background
terms.  The extraction of the 
"bare" verteces
is model-dependent procedure, which is
beyond the scope of this work.

For the resonances $D_{13}(1520)$ and $F_{15}(1680)$,
the  large photoexcitation helicity amplitudes
${}_pA_{3/2}$ are given by the PDG with great accuracy.
By this reason having the contributions of these
resonances to the $\gamma p\rightarrow \eta p$ multipole amplitudes, 
we have the possibility to extract
the branching ratios for 
 $D_{13}(1520)$ and $F_{15}(1680)$
to the $\eta N$ channel
with better accuracy than in the PDG data:
\begin{eqnarray}
&&\Gamma\left(D_{13}(1520)\rightarrow\eta N\right)/
\Gamma\left(D_{13}(1520)\rightarrow all\right)=0.05\pm0.02~\%,\\
&&\Gamma\left(F_{15}(1680)\rightarrow\eta N\right)/
\Gamma\left(F_{15}(1680)\rightarrow all\right)=0.16\pm0.04~\%.
\end{eqnarray}

$\underline{The~g_{\eta NN}~and~g_{\eta NR}~coupling~constants}$.
The values of $g_{\eta NN}$ and $g_{\eta NR}$
found in our analyses are:   $g_{\eta NN}/(4\pi)^{1/2}=0.26$,
$g_{\eta NR}/(4\pi)^{1/2}=-1.12$ in the dispersion relations approach,
and  $g_{\eta NN}/(4\pi)^{1/2}=0.11$,
$g_{\eta NR}/(4\pi)^{1/2}=-0.9$ in the isobar model.
These constants are much smaller than  $g_{\pi NN}$ and  $g_{\pi NR}$:
 $g_{\pi NN}/(4\pi)^{1/2}=3.78$,
$g_{\pi NR}/(4\pi)^{1/2}=-3.1\pm 0.4$.
The last value we have found from
the width of the $P_{11}(1440)\rightarrow \pi N$
decay, the sign is  found
from the pion photoproduction
data.
The small value of $g_{\eta NN}$
was obtained also in the analyses of Refs. \cite{7,13,14}.
It has explanation in the chiral Lagrangian approach \cite{21,22}.
The smallness of the $g_{\eta NN}$ and  $g_{\eta NR}$ coupling constants
leads to the smallness of the $N$
and $P_{11}(1440)$ contributions in comparison
with the dominant $S_{11}(1535)$ contribution.
However, these contributions, as well the $t$-channel
$\rho$ and $\omega$ contributions, which are also
small compared to the $S_{11}(1535)$,
play an important role in getting the better description
of the data with better $\chi^2$.
In Fig. 5 by the thin dotted curves we present the summary
background contributions to the multipole amplitudes,
caused by the Born term, the $P_{11}(1440)$ resonance
and the $\rho$ and $\omega$ $t$-channel exchanges.
These results correspond to the isobar model.
It is seen that they
are very small for all multipole amplitudes,
except $M_{1+}$, where the large background
is generated mostly by the Born term.
As it was mentioned in the Introduction,
dispersion relations allow to find 
the $u$-channel
resonance contributions.
These contributions are presented for the lowest multipoles
in Fig. 5 by the thin dashed curves.
For the multipoles with $l>1$ they are negligibly small.
It is seen that the $u$-channel
resonance contributions are very small for all multipoles,
except $E_{0+}$.

Recently the properties of the $P_{11}(1710)$ resonance
have been discussed in connection with its possible identification
as the member of the baryon anti-decuplet in the chiral solitons
picture \cite{23,24,25}.
If this identification is right, the width of the $P_{11}(1710)$
should be $\simeq~40~MeV$ \cite{23}, and the magnetic moment
of the transition $P_{11}(1710)\rightarrow\gamma N$,
$\mu_{NN^*}\equiv\frac{ \kappa_{NN^*}}{2m_N}$, 
should be in the ranges \cite{24}:
\begin{equation}  
\kappa_{pp^*}=-0.15\div 0.15
~,~\kappa_{nn^*}=-1\div -0.3~.
\end{equation}  
We performed the fit of the experimental data
on the proton, taking $\Gamma\left(P_{11}(1710)\right)
=40~MeV$. It appeared, that the results were
changed only slightly. The magnetic moment 
of the  $P^+_{11}(1710)\rightarrow\gamma p$ transition
found using 
Eqs. (8,9) is equal to
\begin{equation}
\kappa_{pp^*}=0.05\div 0.06~,
\end{equation}
which is within the ranges (30). The preliminary data
on the $\eta$ photoproduction on the neutron obtained
at large angles $120^{\circ}$, $140^{\circ}$, and $160^{\circ}$
show a sharp rise of the ratio 
$\sigma(\gamma n \rightarrow \eta n)/\sigma(\gamma p \rightarrow \eta p)$
in the vicinity of  the $P_{11}(1710)$ \cite{25}.
According to our very approximate estimations,
such behaviour of the cross sections can be described
taking
\begin{equation}
\kappa_{nn^*}=-0.3~.                           
\end{equation}
This means that $|\kappa_{pp^*}|\ll |\kappa_{nn^*}|$,
as it is predicted by the chiral solitons picture.
So, more complete data on  the $\eta$ photoproduction 
on the neutron, which will allow to make more reliable
analysis, may provide very interesting possibility
to find arguments in favour of the
identification of the $P_{11}(1710)$ resonance
as the member of the baryon anti-decuplet. 

In the analysis of Ref. \cite{14} of 
the $\gamma p\rightarrow \eta p$ data the need
for a third $S_{11}$ resonance in the second
resonance region was found.
Such a resonance has been discussed in Ref. \cite{26}.
However, in both our approaches such state
is not supported by the fits to the data.

\section{Conclusion}

In this paper we presented the results of the analysis of  
$\eta$ photoproduction on protons, performed using
data on  cross sections from TAPS \cite{11},
GRAAL \cite{8} and CLAS \cite{9}, on beam asymmetry
from GRAAL \cite{10}, and on target asymmetry 
from ELSA \cite{12}. The analysis is made using both 
fixed-$t$ dispersion relations and an isobar model.
The isobar model we use is very close to that
of Ref. \cite{13}. The difference lies in a different
parametrization of the resonance contributions
and in  our inclusion 
of the  $P_{11}(1440)$ contribution
in the model.
Another difference with the analysis of Ref. \cite{13}
is the inclusion 
of the target asymmetry data 
in our fitting procedure.

The utilization of two approaches
and comparison with the results of Ref. \cite{13}
allow us to clarify the model-dependence of the results obtained.
We conclude that the results for the contributions
of the resonances $D_{13}(1520)$, $S_{11}(1535)$,
$F_{15}(1680)$ to the corresponding multipole amplitudes  
are stable, and do not depend on the model used.
This is connected with the fact that the
$D_{13}(1520)$, $S_{11}(1535)$,
$F_{15}(1680)$ contributions are related directly
to specific features of the cross section and beam asymmetry
data. Determination of the  
$D_{13}(1520)$ and $F_{15}(1680)$ resonance contributions
with good accuracy, in 
combination with the PDG results for the
${}_pA_{3/2}\left(D_{13}(1520)\rightarrow \gamma N\right)$ and
${}_pA_{3/2}\left(F_{15}(1680)\rightarrow \gamma N\right)$ 
helicity amplitudes, allowed us to find the branching ratios for
$D_{13}(1520)\rightarrow \eta N$,
$F_{15}(1680)\rightarrow \eta N$ decays
with significantly better accuracy than the PDG data.

The total Breit-Wigner width
of the  $S_{11}(1535)$ is model-dependent; we have
obtained $\Gamma \left(S_{11}(1520)\right)=142$ and $195~MeV$,
using dispersion relations and the isobar model respectively.
The photoexcitation helicity amplitudes ${}_pA_{1/2}$
for the $S_{11}(1535)\rightarrow p\gamma$
transition obtained in both approaches  
are in good agreement
with the results obtained in other analyses
of the $\eta$ photoproduction data.

The contributions of other resonances
are not related directly to the specific features
of the observables used in the fitting procedure.
For this reason the results for these resonances
are more model-dependent. With this, 
the signs and  orders of  magnitude of
the $S_{11}(1650)$, $D_{15}(1675)$,
$P_{11}(1710)$, $P_{13}(1720)$ contributions
to the corresponding multipole amplitudes
are determined.
The results for the
$D_{13}(1700)$ 
are strongly model-dependent.

The $\eta N$ branching ratios for all  resonances, obtained
combining the PDG  photoexcitation helicity amplitudes 
and the resonance contributions to multipole amplitudes
from our analysis, are within the ranges given by the PDG.
The only exception is the $P_{13}(1720)$
resonance. 
Here we had to take larger $\beta_{\eta N}=18-25\%$ in order 
to stay within the ranges for ${}_pA_{1/2}$ given by the PDG.

\begin{center}
{\large {\bf {Acknowledgments}}}
\end{center}

I am thankful to V. D. Burkert, who has initiated my interest
to this investigation. I am grateful to L. Tiator
for very useful correspondence, to S. Dytman
and V. Mokeev for useful discissions, and to M. Dugger and
Wen-Tai Chiang for their help in getting data.
I express my gratitude for the hospitality at Jefferson Lab
where the last stage of this work was accomplished.
 \vspace{1cm}

{\Large \bf Appendix A. Relations between invariant 
and multipole amplitudes}
\vspace{0.3cm}
\renewcommand\theequation{A.\arabic{equation}}
\setcounter{equation} 0

In order to connect invariant and multipole amplitudes
it is convenient to introduce
the intermediate amplitudes $f_i(s,cos \theta,Q^2)$: 
\begin{eqnarray}
f_1=&&\left[(W-m_N)B_1-B_5\right]\frac
{\left[(E_1+m_N)(E_2+m_N)\right]^{1/2}}{8\pi W},\nonumber \\
f_2=&&\left[-(W+m_N)B_1-B_5\right]\frac
{\left[(E_1-m_N)(E_2-m_N)\right]^{1/2}}{8\pi W}, \\
f_3=&&\left[2B_3-B_2+(W+m_N)\left(\frac{B_6}{2}-B_8 \right)\right]
\frac{\left[(E_1-m_N)(E_2-m_N)\right]^{1/2} 
(E_2+m)}{8\pi W},\nonumber \\
f_4=&&\left[-(2B_3-B_2)+(W-m_N)\left(\frac{B_6}{2}-B_8\right)\right]
\frac{\left[(E_1+m_N)(E_2+m_N)\right]
^{1/2}(E_2-m)}{8\pi W},\nonumber\\
f_5=&&\left\{\left[Q^2B_1+(W-m_N)B_5
+2W(E_1-m_N)\left(B_2-\frac{W+m_N}{2}B_6
\right)\right](E_1+m_N)\right.\nonumber \\
&&\left.-X\left[(2B_3-B_2)+(W+m_N)
\left(\frac{B_6}{2}- B_8 \right)\right]\right\}
\frac{(E_1-m_N)(E_2+m_N)}{8\pi WQ^2},\nonumber \\
f_6=&&\left\{-\left[Q^2B_1-(W+m_N)B_5
+2W(E_1+m_N)\left(B_2+\frac{W-m_N}{2}B_6
\right)\right](E_1-m_N)\right.\nonumber \\
&&\left.+X\left[(2B_3-B_2)-(W-m_N)
\left(\frac{B_6}{2}- B_8 \right) \right]\right\}
\frac{(E_1+m_N)(E_2-m_N)}{8\pi WQ^2},\nonumber
\end{eqnarray}
where
\begin{equation}
X=\frac{\tilde{k}_0}{2}(t-m_{\eta} ^2+Q^2)-Q^2\tilde{q}_0,
\end{equation}
$\theta$ is the polar angle of $\eta$ in the c.m.s.,
$\tilde{k}_0,\tilde{q}_0,E_1,E_2$ are the energies
of the virtual photon, eta, initial and final nucleons
in this system.

The expansion of the intermediate amplitudes over multipole
amplitudes $M_{l\pm}(s,Q^2)$, $E_{l\pm}(s,Q^2)$, 
$S_{l\pm}(s,Q^2)$
has the form:
\begin{eqnarray}
&&f_1=\sum\left\{(lM_{l+}+E_{l+})P'_{l+1}(cos\theta)+
\left[(l+1)M_{l-}+E_{l-} \right]
P'_{l-1}(cos\theta)\right\},\nonumber \\
&&f_2=\sum\left[(l+1)M_{l+}+lM_{l-} \right] P'_l(cos\theta),\nonumber \\
&&f_3=\sum\left[(E_{l+}-M_{l+})P''_{l+1}(cos\theta)+
(E_{l-}+M_{l-})P''_{l-1}(cos\theta)\right],\\
&&f_4=\sum(M_{l+}-E_{l+}-M_{l-}-E_{l-})P''_l(cos\theta),\nonumber \\
&&f_5=\sum\left[(l+1)S_{l+}P'_{l+1}(cos\theta)-
lS_{l-}P'_{l-1}(cos\theta)\right],\nonumber \\
&&f_6=\sum\left[lS_{l-}-(l+1)S_{l+}\right] P'_l(cos\theta).\nonumber
\end{eqnarray}
The amplitudes 
$f_i(s,cos \theta,Q^2)$ are related to the helicity
amplitudes by:
\begin{eqnarray}
&&H_1=-\cos\frac{\theta }{2} \sin \theta
(f_3+f_4)/\sqrt{2},\nonumber \\
&&H_2 =-\sqrt{2} \cos\frac{\theta }{2}
\left[f_1-f_2- \sin^2\frac{\theta }{2}(f_3-f_4)\right],\nonumber\\
&&H_3=\sin\frac{\theta }{2} \sin \theta
(f_3-f_4)/\sqrt{2},\\
&&H_4 =\sqrt{2} \sin\frac{\theta }{2}
\left[f_1+f_2+ \cos^2\frac{\theta }{2}(f_3+f_4)\right]
,\nonumber \\
&&H_5 =-\frac{Q}{k} \cos
\frac{\theta }{2}(f_5+f_6),\nonumber \\
&&H_6 =\frac{Q}{k} \sin
\frac{\theta }{2}(f_5-f_6).\nonumber
\end{eqnarray}
In the case of the photoproduction, the differential cross section,
the polarization of the final nucleon, and 
the polarized beam and target
asymmetries are related to the helicity amplitudes
in the following way:
\begin{eqnarray}
&&\sigma\equiv\frac{d\sigma}{d\Omega}=\frac{1}{2}\frac{q}{k}(|H_1|^2+
|H_2|^2+|H_3|^2+|H_4|^2),\nonumber \\
&&P=-\frac{q}{k}\frac{1}{\sigma}{\cal I}m(H_1H_3^*+H_2H_4^*),\\ 
&&\Sigma=\frac{q}{k}\frac{1}{\sigma}{\cal R}e(H_1H_4^*-H_2H_3^*),\nonumber\\ 
&&T=\frac{q}{k}\frac{1}{\sigma}{\cal I}m(H_1H_2^*+H_3H_4^*).\nonumber 
\end{eqnarray}

\newpage
{\Large \bf {Figure Captions}}
\vspace{1cm}
\newline
{\large \bf{Fig. 1}}. 
Differential cross section for $\gamma p\rightarrow \eta p$.
The solid and dashed curves are the results obtained using
dispersion relations and isobar model.
The data are from TAPS \cite{11} (open circles),
GRAAL \cite{8} (full squares), and CLAS \cite{9} (full triangles).
\newline
{\large \bf{Fig. 2}}.  
Polarized beam asymmetry  for $\gamma p\rightarrow \eta p$.
The solid and dashed curves are the results obtained using
dispersion relations and isobar model. The dotted curve
at $E_{\gamma}=1050~MeV$ are the results
of the isobar model with the $F_{15}(1680)$ contribution
taken to be equal 0.
The data are from GRAAL \cite{10}.
\newline
{\large \bf{Fig. 3}}.  
Polarized target asymmetry  for $\gamma p\rightarrow \eta p$.
The solid and dashed curves are the results obtained using
dispersion relations and isobar model.
The data are from ELSA \cite{12}.
\newline
{\large \bf{Fig. 4}}. 
Total cross section for $\gamma p\rightarrow \eta p$.
The solid and dotted curves are the results obtained using
dispersion relations and isobar model.
The data are from TAPS \cite{11} (open circles),
GRAAL \cite{8} (full circles), and  CLAS \cite{9} (open squares).
\newline
{\large \bf{Fig. 5}}. The results for the 
multipole amplitudes. 
Solid (dotted) and dashed (dash-dotted) curves are the real and imaginary
parts of the amplitudes obtained using dispersion relations (isobar model).
By the thin dotted curves the isobar model
background contributions 
caused by the Born term, the $P_{11}(1440)$ resonance
and the $\rho$ and $\omega$ $t$-channel exchanges are presented.
By the thin dashed curves
the $u$-channel
resonance contributions found using
dispersion relations 
are presented.
\vspace{1cm}
\newline
{\Large \bf {Table Captions}}
\vspace{1cm}
\newline
{\large \bf{Table 2}}. 
Masses, widths, and branching ratios of the resonances.
In the first rows the ranges
given by the PDG \cite{17} are presented. In the second
rows we list the widths used in our analysis.
The masses and widths marked by the stars are found from the
fitting of the data: on the second rows using
dispersion relations and on the third rows
using the isobar model. The branching ratios $\beta_{\eta N}$
in the second and third rows are used for the calculation of the
photoexcitation helicity amplitudes presented in Table 3.
\newline
{\large \bf{Table 3}}. 
Resonance contributions into the multipole amplitudes at the resonance
positions (in $\mu b^{1/2}$) and
photoexcitation helicity amplitudes (in $10^{-3}~GeV^{-1/2}$).
The values on the first and second rows are found in this work
using dispersion relations and isobar model, respectively. The values on the third
rows for the multipole amplitudes are from Ref. \cite{13};
for the photoexcitation helicity amplitudes
on the third rows
the PDG data \cite{17} are presented.
$\zeta_{\eta N}$
are the relative signs between the $NN\eta$ and  $N^*N\eta$ coupling constants
which are used for the calculation of the
photoexcitation helicity amplitudes by the relation (23).

\newpage
\begin{table}[t]
\begin{center}
\begin{tabular}{cccccc}
\hline
$V$&&$m_V~(MeV)$&${\tilde g}_1$&${\tilde g}_2/{\tilde g}_1$&$\lambda_V$\\
\hline
$\rho$&&768.5&1.8-3.2&4.3-6.6&$1.02\pm 0.09$\\
&&&1.6&5.4&\\
$\omega$&&782.6&8-14&0-(-1)&$0.29\pm .03$\\
&&&14.7&-0.55&\\
\hline
\end{tabular}
\caption{\label{tab1}Parameters for the vector mesons.
The ranges for ${\tilde{g}}_{1,2}$ are from Ref. \cite{19},
the values of $\lambda_V$
are extracted from the PDG data \cite {17} for
$V\rightarrow \eta\gamma$.
The values on the second rows are obtained in this work
from the analysis of the  $\gamma p\rightarrow \eta p$
data using the isobar model.}
\end{center}
\end{table}
\newpage
\begin{table}[t]
\begin{center}
\begin{tabular}{cccccc}
\hline
$Resonance$&&$M~(MeV)$&$\Gamma~(MeV)$&$\beta_{\eta N}~(\%)$&$\beta_{\pi N}~(\%)$\\
\hline
$D_{13}(1520)$&&1515-1530&110-135&$0\pm1$&50-60\\
&&$1510^*$&120&0.032&\\
&&$1523^*$&${}$&0.073&${}$\\
$S_{11}(1535)$&&1520-1555&100-200&30-55&35-55\\
&&$1527^*$&$142^*$&50&\\
&&$1542^*$&$195^*$&50&\\
$S_{11}(1650)$&&1640-1680&145-190&3-10&55-90\\
&&$1648^*$&$85^*$&3.8&\\
&&$1649^*$&$125^*$&5.5&\\
$D_{15}(1675)$&&1670-1685&140-180&$0\pm1$&40-50\\
&&$1666^*$&160&1.2&\\
&&$1663^*$&${}$&1.4&${}$\\
$F_{15}(1680)$&&1675-1690&120-140&$0\pm1$&60-70\\
&&$1670^*$&130&0.13&\\
&&$1676^*$&${}$&0.18&${}$\\
$D_{13}(1700)$&&1650-1750&50-150&$0\pm1$&5-15\\
&&$1634^*$&100&1&\\
&&$1727^*$&${}$&1&${}$\\
$P_{11}(1710)$&&1680-1740&50-250&$6\pm1$&10-20\\
&&$1710^*$&100&7&\\
&&$1740^*$&${}$&6&${}$\\
$P_{13}(1720)$&&1650-1750&100-200&$4\pm1$&10-20\\
&&$1720^*$&150&25&\\
&&$1750^*$&${}$&18&${}$\\
\hline
\end{tabular}
\caption{\label{tab2}}
\end{center}
\end{table}

\begin{table}[t]
\begin{center}
\begin{tabular}{cccccccccccccc}
\hline
Resonance&&$E_{l\pm}$&&&$M_{l\pm}$&&&${}_pA_{1/2}$&&&${}_pA_{3/2}$
&$\zeta_{\eta N}$\\
\hline
$D_{13}(1520)$&&0.053&&&0.029&&&-40&&&166&+1\\
&&0.071&&&0.040&&&-43&&&166&+1\\
&&0.062&&&0.038&&&$-24\pm9$&&&$166\pm5$\\
$S_{11}(1535)$&&1.853&&&&&&96&&&&+1\\
&&1.832&&&&&&119&&&&+1\\
&&1.845&&&&&&$90\pm30$&&&\\
$S_{11}(1650)$&&-0.268&&&&&&53&&&&-1\\
&&-0.411&&&&&&53&&&&-1\\
&&-0.437&&&&&&$53\pm16$&&&\\
$D_{15}(1675)$&&-0.012&&&-0.008&&&27&&&-4&-1\\
&&-0.011&&&-0.012&&&27&&&1&-1\\
&&0&&&0.078&&&$19\pm8$&&&$15\pm9$\\
$F_{15}(1680)$&&0.024&&&0.015&&&-15&&&133&+1\\
&&0.029&&&0.018&&&-14&&&133&+1\\
&&0.015&&&0.01&&&$-14\pm6$&&&$133\pm12$\\
$D_{13}(1700)$&&0.015&&&-0.029&&&-30&&&7&-1\\
&&0.001&&&-0.001&&&-1&&&0&-1\\
&&0.009&&&-0.007&&&$-18\pm13$&&&$-2\pm24$\\
$P_{11}(1710)$&&&&&0.172&&&29&&&&-1\\
&&&&&0.127&&&24&&&&-1\\
&&&&&-0.26&&&$9\pm22$&&&\\
$P_{13}(1720)$&&0.138&&&0.199&&&48&&&8&+1\\
&&0.126&&&0.124&&&48&&&0&+1\\
&&0.032&&&-0.016&&&$18\pm30$&&&$-19\pm20$\\
\hline
\end{tabular}
\caption{\label{tab3}}
\end{center}
\end{table}
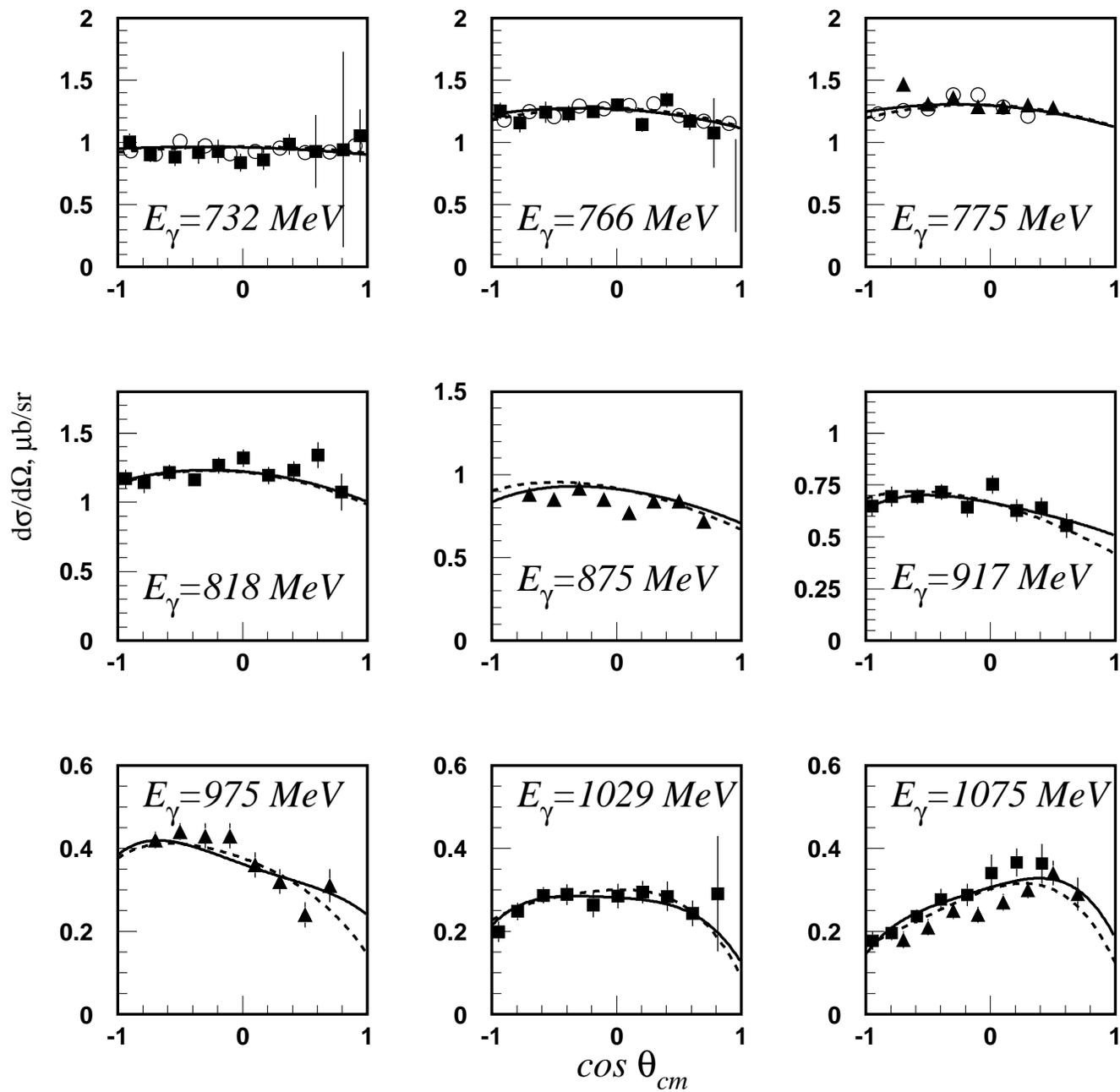
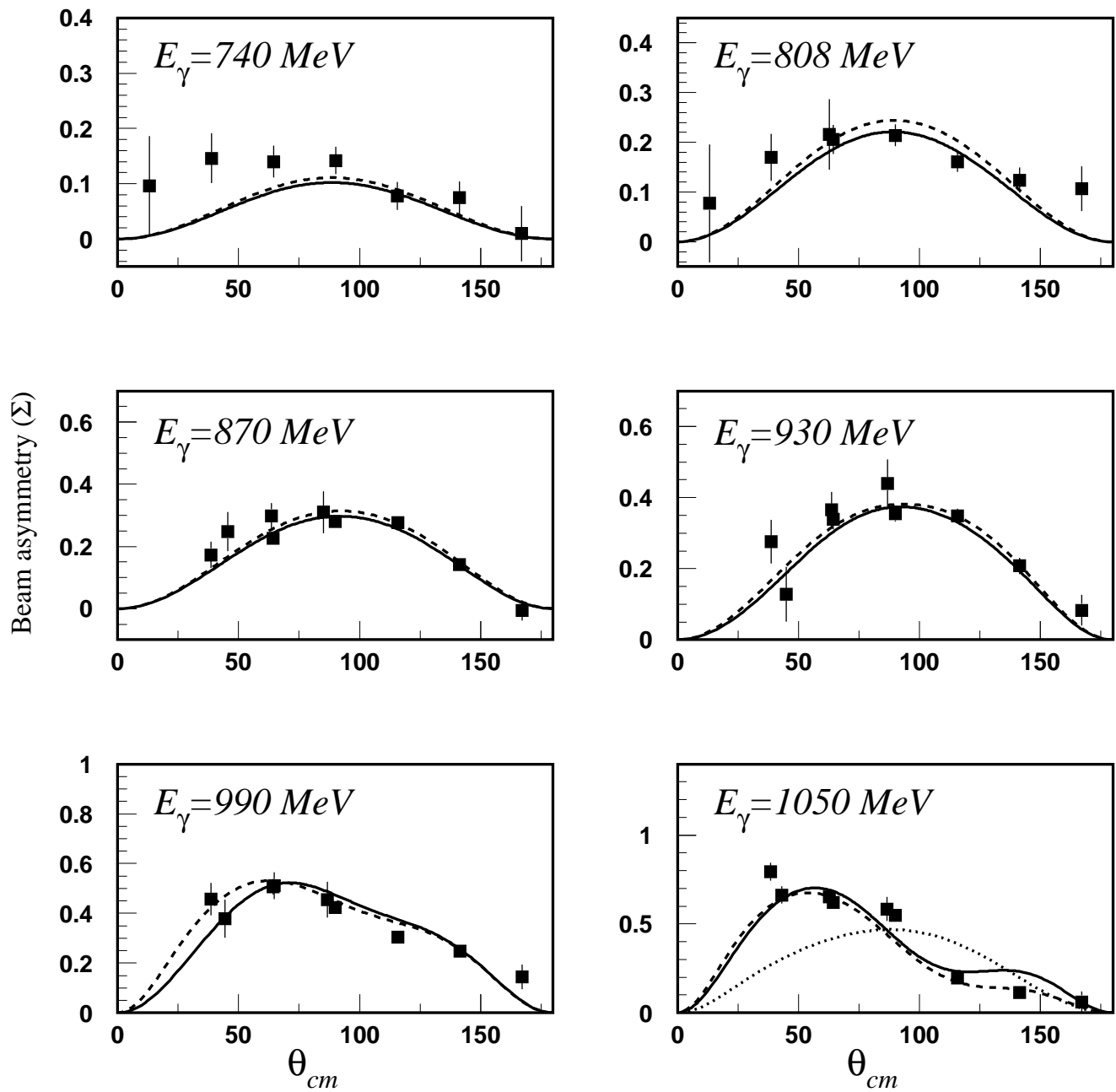
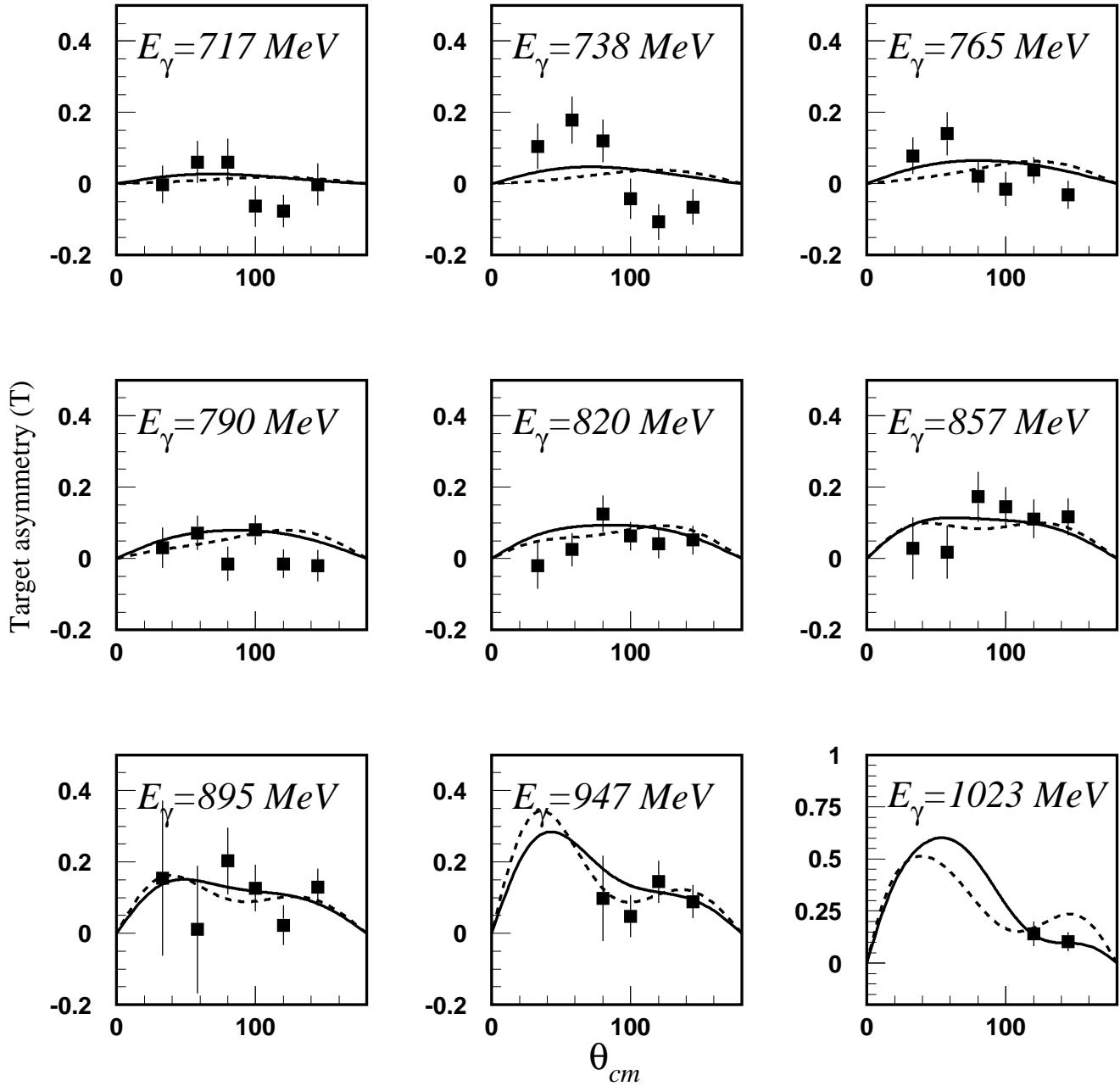
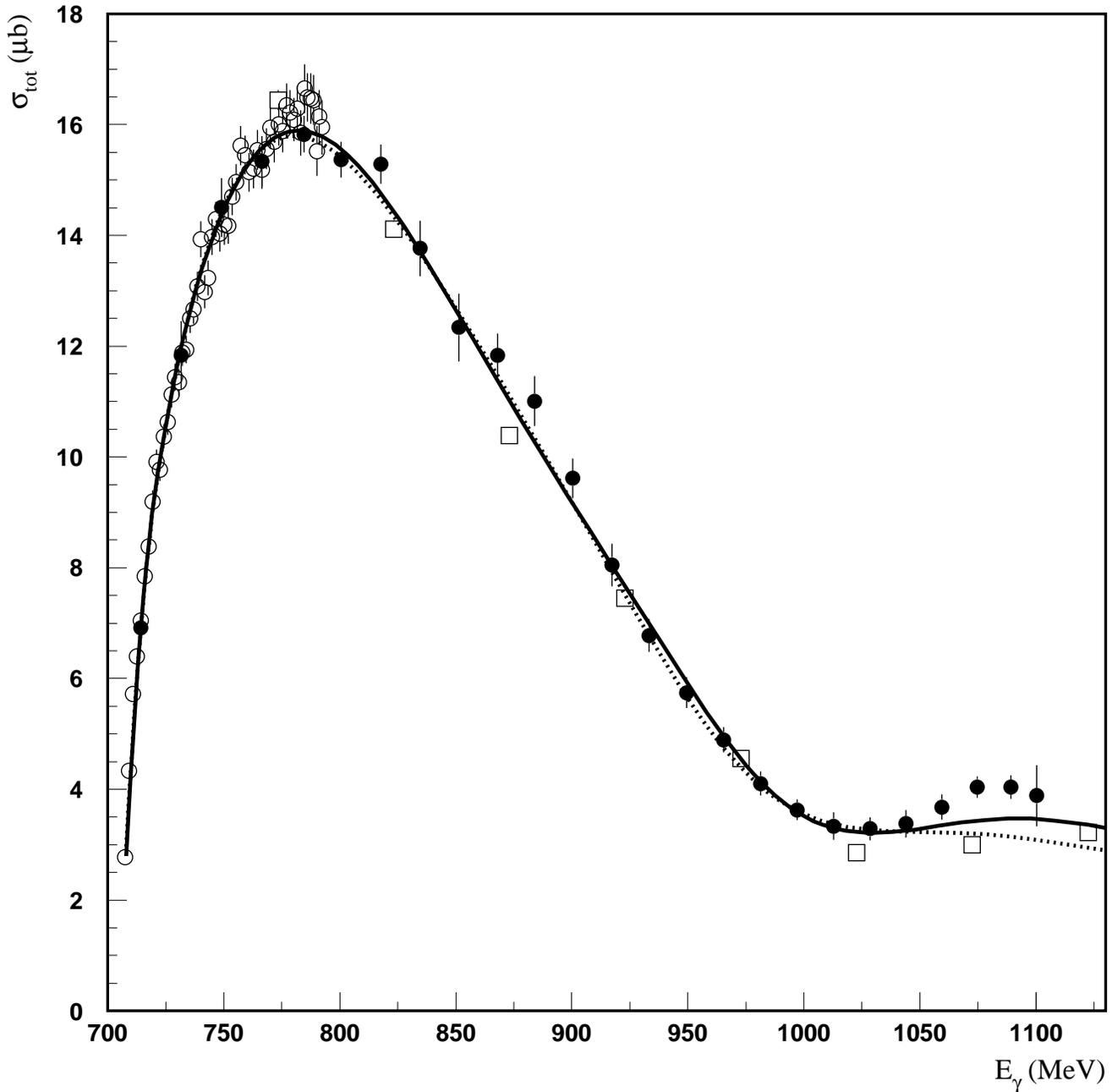
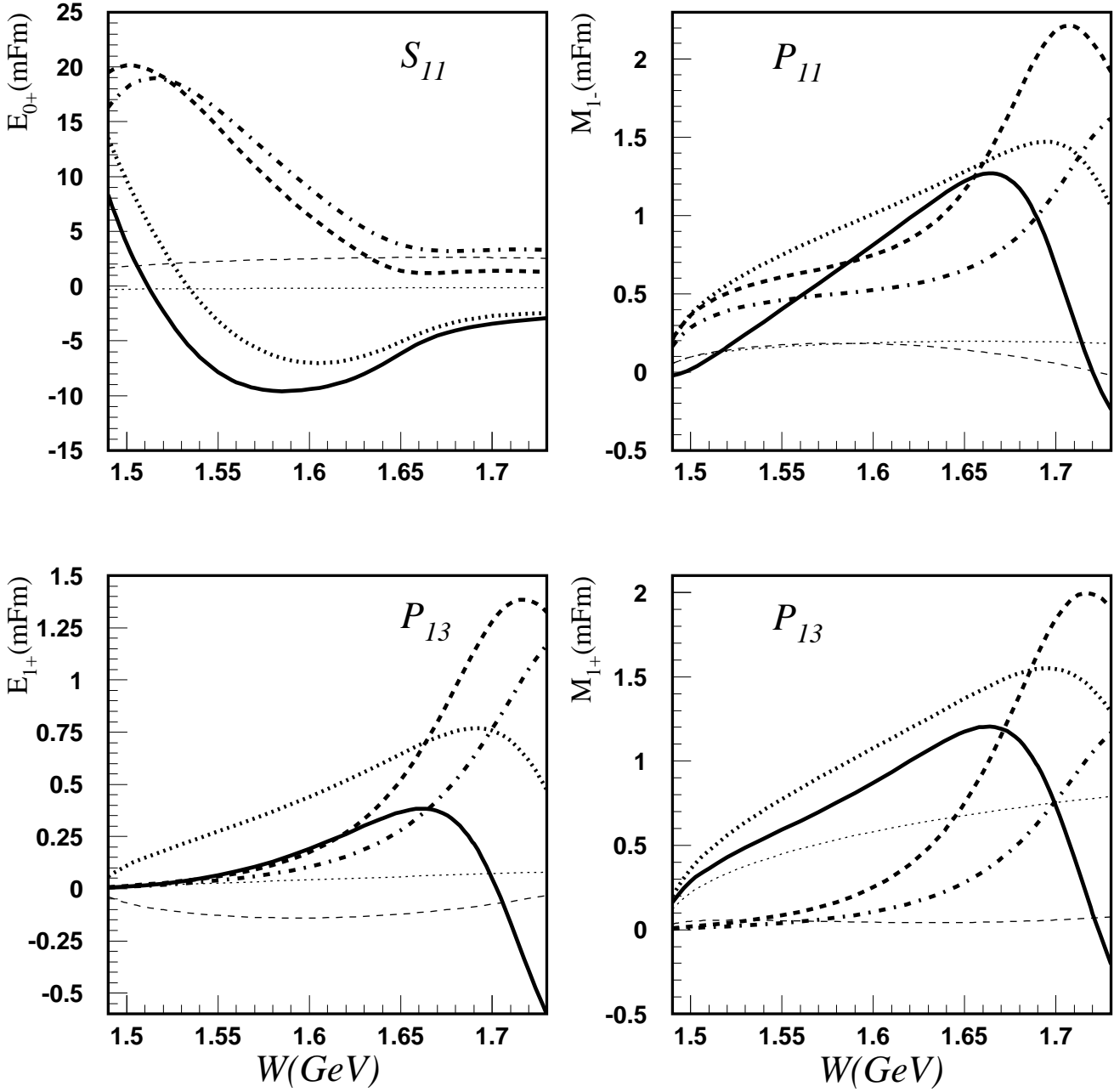
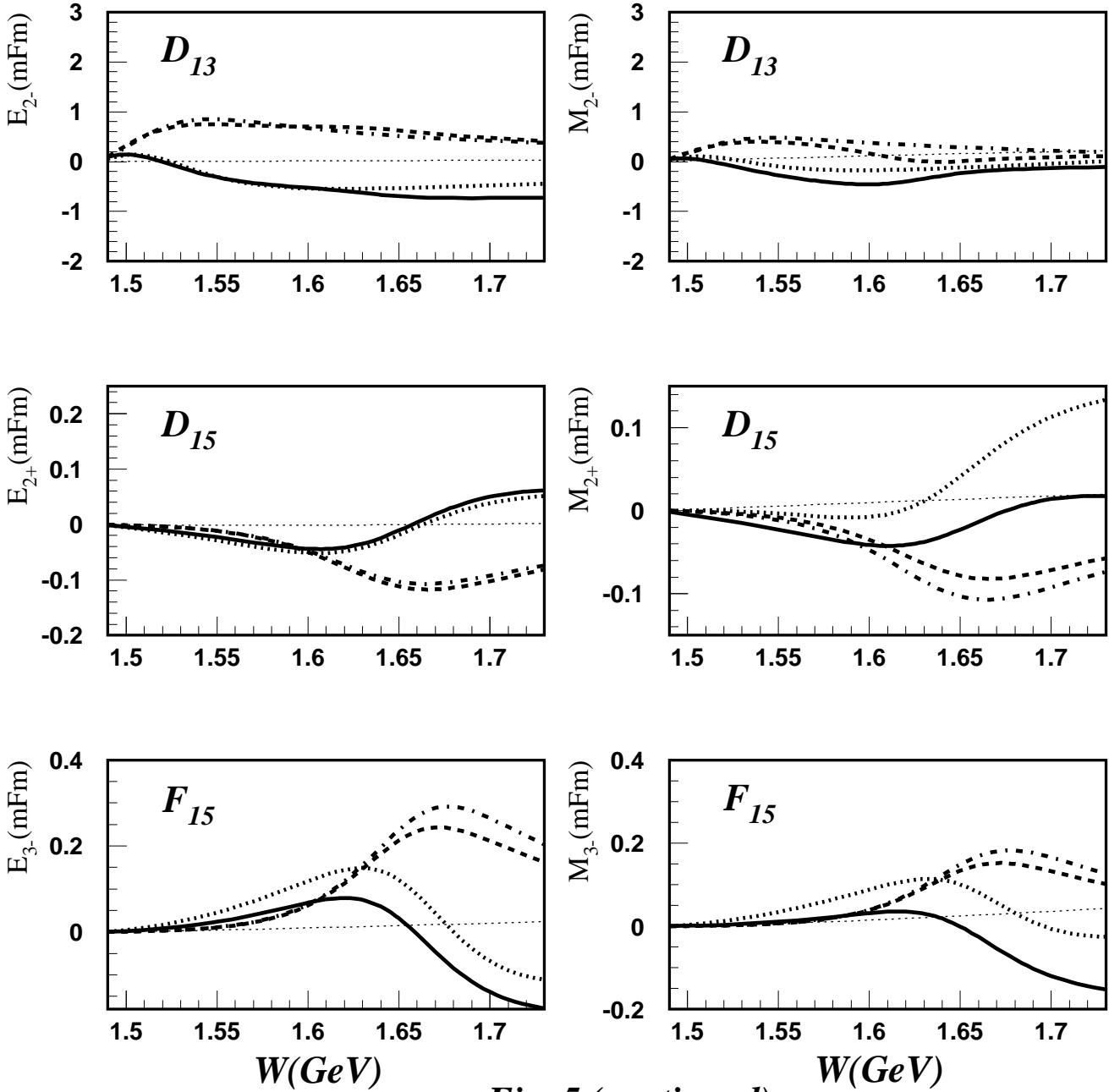
\end{document}